\documentclass[twocolumn,showpacs,aps,prb]{revtex4}
\usepackage{dcolumn} 
\usepackage{graphicx}
\usepackage{subfigure}
\usepackage{epsfig}
\usepackage{epstopdf}
\begin{document}
\title{Conservation of  dielectric constant upon  amorphization in  perovskite oxides}

\author{Pietro Delugas,$^1$ Vincenzo Fiorentini,$^{1,2}$
Alessio Filippetti,$^2$ and Geoffrey Pourtois$^3$}

\affiliation{1) NXP Semiconductors, Kapeldreef 75, B-3001 Leuven, Belgium\\ 2) SLACS-CNR-INFM (Sardinian LAboratory for Computational
materials Science) and Dipartimento di  Fisica, Universit\`a di Cagliari,
Cittadella  Universitaria, I-09042 Monserrato (CA), Italy\\
3) IMEC, Kapeldreef 75, B-3001 Leuven, Belgium}
\date{}
\begin{abstract} 
We report calculations indicating that  amorphous RAO$_3$  oxides, with R and A trivalent cations, have  approximately the same static dielectric constant  as their perovskite crystal phase. The effect  is due to the disorder-activated polar response of    non-polar  crystal modes at low frequency, which compensates a moderate but appreciable reduction of the   ionic dynamical charges.  The dielectric response was studied via density-functional perturbation theory. Amorphous samples were generated by  molecular dynamics melt-and-quench simulations.
\end{abstract} 
\pacs{77.22-d,63.20-e,78.30-j,61.66-f}

\maketitle
\section{Introduction}
In the context of integrated microelectronics there is a clear dichotomy, rooted in  the fundamentals of  dielectric screening, between  crystalline and   amorphous dielectric layers. The latter are preferable in terms of electrical leakage behavior, but their  dielectric response may be unsatisfactory for applications requiring a large static dielectric constant ${\kappa}_s$.  A case in point is that of the so-called high-$\kappa$ materials, gradually replacing silica \cite{hik} as gate dielectric in aggressively downscaled Si-based integrated microelectronics.  It  is unclear in general if the  high $\kappa$ of a crystal compound will be conserved  in its amorphous phase.  This is due to the nature of the dielectric screening  in high-$\kappa$ materials, where the total static dielectric constant $\kappa_s$=${\kappa}_{\infty}$+$\kappa_{\rm ion}$ is dominated \cite{lalo,rev,dysco} by the 
 lattice-vibrational  component ${\kappa}_{\rm ion}$, being typically a factor 2 to 10 larger than  the electronic ${\kappa}_{\infty}$. Large values of ${\kappa}_{\rm ion}$ are due  to large  effective charges and   soft infrared (IR)  vibrational modes. This is borne out by   its definition: \begin{equation}
\kappa^{\alpha\beta}_{\rm ion} = \frac{4\pi e^2}{\Omega}
\sum_{\lambda} 
\frac{z_{\lambda\alpha} z_{\lambda\beta}}{\omega^2_{\lambda}},
\ \ \   
z_{\lambda\beta}=\sum_{i\beta} \frac{Z^*_{i,\alpha\beta}
 \, \xi_{i,\lambda\beta}}{\sqrt{M_i}},
\end{equation}
with  $z$  the mode charge vector, $\Omega$ the system volume, ${Z}^*_{i,\alpha,\beta}$ the effective or dynamical or Born charge tensor 
and $M_i$ the mass of atom $i$, 
$\xi_{i,\lambda\beta}$ and $\omega$ the  eigenvector and eigenfrequency of mode
$\lambda$ at zero wave-vector, and $\alpha$, $\beta$ cartesian indexes. 

Dynamical charges are termed "anomalous" if they  exceed the nominal ionicity. Such "anomaly" (sort of a misnomer, being ubiquitous and dramatic in many oxides) results from a subtle interplay of covalent and ionic bonding, as discussed in detail in Ref.\,\onlinecite{zstar}. Lattice disorder will generally reduce the efficiency of orbital overlap and, hence, the dynamical charge flow measured by the dynamical charges. Thus the  $Z^*$'s  are  expected to decrease in  the disordered amorphous environment. On the other hand,  softer  IR-active modes counterbalancing this effect (through the 1/$\omega^2$ factor in Eq.\,1) may appear in the amorphous phase (they indeed do, both in the present case as well as in zirconia, as reported in Ref.\,\onlinecite{zhao-zro2}). The major 
reason for this is that IR-active and inactive modes are not discriminated by symmetry constraints as in the crystal phase, so that  non-IR crystal modes, or remnants thereof, can acquire IR activity. For sufficiently low frequency, thanks to the 1/$\omega^2$ factor in Eq.\,1, even weakly IR-active modes may produce  a large dielectric intensity (counterbalancing the general  decrease in $Z^*$) and cause a conservation or a boost of the dielectric constant. 

 From the discussion so far, it is apparent that a quantitative assessment of the dielectric constant ingredients  is   central to evaluating a technologically relevant dielectric. Indeed, if the amorphous phase of a given material has a similar ionic screening as its crystal phase, there is hardly any point in pursuing the fabrication of single-crystal layer of that material.

 Here we study from first principles the static dielectric response  in amorphous high-$\kappa$  RAO$_3$ oxides with A,   R trivalent cations. We show that indeed   the dielectric constant  in the amorphous phase is  similar to that of their distorted-perovskite crystal phase, due to
a moderate  reduction in polarizability, and to the disorder-induced IR activation of non-polar low-energy modes of the crystal.

Our specific case studies are   LaAlO$_3$ and DyScO$_3$. We described previously their  crystal properties.\cite{lalo,dysco}  The nomenclature "R" for the larger, nominally 12-fold coordinated  cation and "A" for the octahedrally-coordinated one is used throughout the paper. These two compounds  represent quite thoroughly the relevant trivalent-cation perovskites RAO$_3$ for high-$\kappa$ applications. Indeed, the  choice of A can be restricted to either  Al or Sc because rare-earths, Lu, La, and Y  function as the R cation,  mixed-valent Ti and Mn  cause  small gaps  and   correlation,\cite{latio} and Ga is analogous to Al in this stoichiometry.\cite{lalo,lagao}  Both materials  are actively studied by the high-$\kappa$ community, and both have been reported \cite{laloamo,discoamo} to  have high  dielectric constants ($\sim$25)  in the  crystalline  {\it as well as} the amorphous state. In DyScO$_3$ the effect  is so far  uncontroversial. A report \cite{lalolow} also exists of a sizable (40\%) deterioration of  the dielectric constant  in LaAlO$_3$.  While not aiming at a definite resolution of this issue, here we  find   that a conservation of dielectric screening in these materials is plausible from the theoretical point of view.

\section{Technical matters}
Amorphous LaAlO$_3$ and DyScO$_3$ samples were generated by a melt-and-quench  molecular dynamics procedure, using the empirical-potential GULP code.\cite{gulp} A liquid was equilibrated at temperature T=5000 K for 30 ps to achieve fully randomized  cation positions. It was then quenched  by  Nose constant-T dynamics at 50 K/ps.  2-ps contant-T intervals separated by abrupt 100 K down-steps were used down to 2000 K. After that, since the mobility was then very low,  larger down-steps of 500 K and simulation periods up to 5 ps were used down to 500 K, whereby structural properties (expected to be well converged\cite{cool}  at this cooling rate)  were sampled in a 5 ps run. 

For LaAlO$_3$, we also ran  ab initio density-functional constant-T dynamics with the VASP code \cite{vasp} to quench the equilibrated liquid from 5000  to 500 K with  alternating 0.5-ps  constant-T intervals and  0.5 ps, 500 K  ramp-down segments. The differences in structural properties in the two approaches are moderate  and   affect  the dielectric properties slightly (see the discussion below). Cell sizes (80 atoms for ab initio dynamics, 80 to 320 atoms for empirical) do not  influence the structure appreciably. 

After cooling to 0 K by damped dynamics,  we further relaxed  the structure using the ab initio Espresso \cite{espresso} code (finding, incidentally, no major changes in structure or  density). On the final structure,  the Espresso code was used to perform linear-response density-functional perturbation theory \cite{dfpt} calculations of electronic dielectric constant,  effective charges and  k=0 phonon modes,  building up the dielectric constant (Eq.\,1).  All DFT calculations are done within the generalized-gradient approximation  (for details see \onlinecite{vasp,espresso}), and will  be indicated henceforth  by the acronym GGA.

\begin{figure}[ht]
\centering
\includegraphics[clip,width=8.5cm]{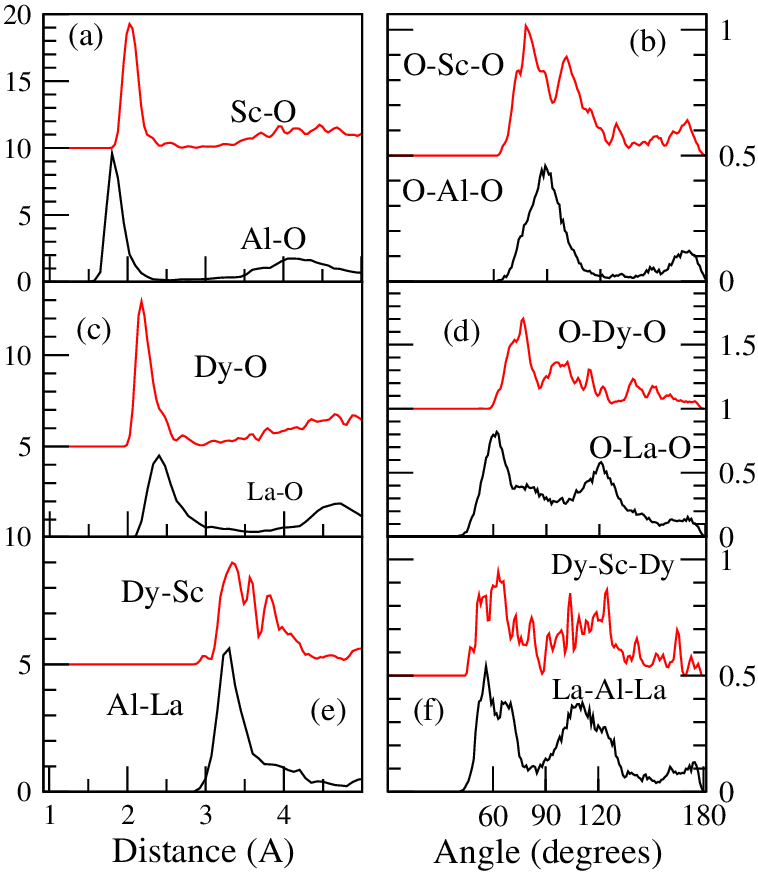}
\caption{\small{Pair correlation functions g(r) (left column) and angle distributions (right column) for amorphous LaAlO$_3$ and DyScO$_3$. The g(r) functions are: (a), A-O; (c), R-O;  (e), R-A. The angle distributions are: (b), O-A-O; (d), O-R-O; (f) R-A-R. The functions for a-DyScO$_3$ are offset vertically for clarity.}} \label{fig1}
\end{figure}

\section{Results and discussion}
\subsection{Structural properties} As no experimental structural data are available for the amorphous materials under consideration, a sanity check of our generated structures is to compare their properties with those of the crystal phase. The amorphous samples appear to  inherit many structural characteristics of the crystal.\cite{lalo,dysco} Their main feature is the approximate conservation of an octahedral A-O backbone, though with somewhat different  octahedra  distortion and degrees of cation disorder  in LaAlO$_3$  and  DyScO$_3$. The structure-characterizing correlation function of the simulated amorphous samples are collected in Fig.\ref{fig1}. 
Preliminarly  we note that  the intermixing of  cation sublattices (i.e A$_{\rm R}$ antisites and such)   is ruled out: no R$\rightarrow$A cation exchange is observed in any of the simulations, and the hypothetical AlLaO$_3$ perovskite is  metallic, 40\% less dense  and widely higher in  energy than LaAlO$_3$. 

Let us discuss Fig.\ref{fig1}. We refer to  crystal or amorphous  phases  by the prefixes "c-" or "a-".  In ideal perovskite the R (La, Dy) cation is 12-fold, and the A (Al, Sc) cation is 6-fold coordinated. The main feature of both a-phases is the A-O g(r) function  peaking  neatly at typical A-O distances (Fig.\ref{fig1}a) with a resulting A-O coordination of 6 as in the c-phases.  The typical O-A-O (e.g. O-Al-O) angles are very close 90 and 180$^{\circ}$ in the crystals; as seen in Fig.\ref{fig1}b, in a-LaAlO$_3$ the angle distribution is a broad peak centered at 90$^{\circ}$, whereas in a-DyScO$_3$ the same function is bimodal with peaks at about 80 and 100$^{\circ}$. This is due to the amplification of a Jahn-Teller-like asymmetry of the octahedron already incipient in the crystal.\cite{dysco}. 
The a-phases  can thus be characterized synthetically  as disordered networks of apex-sharing octahedra.

The R-cations La and Dy are both approximately 6-fold coordinated in the respective a-phases. This means a severe reduction for La, whose coordination  in c-LaAlO$_3$ is close to ideal,\cite{lalo} but almost no change for  Dy which  already had  \cite{dysco} a low coordination of 6 in c-DyScO$_3$ due to the large orthorhombic rotations. The Dy-O g(r) is also sharper than the La-O (Fig.\ref{fig1}c), suggesting a lesser degree of disorder of the Dy sublattice. The loss of coordination of La is also signaled by the O-La-O angle distribution (Fig.\ref{fig1}d), whose peaks at 60 and 120$^{\circ}$ are quite broadened but well-defined, whereas those at  90$^{\circ}$ and 180$^{\circ}$ are both washed out. This behavior is not unexpected as the ideal-perovskite angles are those  internal to a  cuboctahedron (60, 90, 120, and 180$^{\circ}$): the 60$^{\circ}$ is the most robust as it  corresponds to O pairs belonging to one single octahedron, and correlates with the 120$^{\circ}$, as does the 90$^{\circ}$ with the  180$^{\circ}$ one.

Finally, the R-A functions show significant  remnants of the correlation of the interpenetrating cubic sublattices of A and R cations in the c-phase: the  g(r) main peak (Fig.\ref{fig1}e) is fairly well defined at the typical R-A distances, and the R-A-R angle function (Fig.\ref{fig1}f) still has a bimodal structure around the ideal-perovskite angles 70 and 110$^{\circ}$.
 
\subsection{Polarizability} With reference to Eq.\,1, we note that the dipoles $z$ generated by atomic vibrations determine, along with the vibrational frequencies discussed below, the static dielectric constant (Eq.1). One key  ingredient of these dipoles is the ionic dynamical polarizability, i.e. as the time-integrated current flowing upon unit displacement  of a specific ion.\cite{zstar} An unbiased  measure thereof  is provided by the effective dynamical-charge tensor of atom $i$,
$Z^*_{i,\alpha\beta}$=$\Omega$/e ($\partial P_{\alpha}$/$\partial u_{\beta}$), i.e. the  derivative of the total polarization vector with respect to the vector  displacement of atom $i$.\cite{zstar} A drop  in  dynamical charges implies therefore a reduction of  the ionic dielectric activity (Eq.1). 

We compare the present GGA dynamical charges, electronic dielectric constants, and total static dielectric constants with their crystal counterparts \cite{lalo,dysco} in Table \ref{tab1}.  To present the data    synthetically,  we average  the a-phase charge tensors over  all atoms of a given species, and decompose them thereafter into $s$, $p$, and $d$ components (i.e.   spherical, traceless symmetric, and traceless antisymmetric). The $s$ component,  accounting for about 98\% and 95\% of the tensor norm for cations and oxygen respectively, is reported  in Table \ref{tab1} for the amorphous case. In the same Table we also report the $s$ component of  the dielectric tensors (these are only very mildly non-isotropic). As we now discuss, the  dynamical charges results of Table \ref{tab1} are  quite coherent with the  structural properties described above, and with the general properties of the cations involved; charges on O largely adjust to obey dynamical charge-neutrality sum rules.
 
\begin{table}[htb]
  \centering
 \caption{Average effective-charge, electronic, and static dielectric constant tensors in amorphous DyScO$_3$ and LaAlO$_3$, compared to crystal values. See text for details.}

\begin{tabular}{cccc|cc}
\hline\hline    
\multicolumn{1}{c}{DyScO$_3$}&
\multicolumn{1}{c}{$<$$Z_{\rm Sc}^*$$>$}&
\multicolumn{1}{c}{$<$$Z_{\rm Dy}^*$$>$ }&
\multicolumn{1}{c}{$<$$Z_{\rm O}^*$$>$}&
\multicolumn{1}{c}{$\kappa_{\infty}$}&
\multicolumn{1}{c}{$\kappa_{s}$}\\
    \hline 
  crystal       & 4.00 &    3.82 &  --2.61 & 4.8 & 24.0 \\
 amorphous     & 3.63 &    3.83 &  --2.43 & 4.5 & 22.1 \\
 \hline\hline
\multicolumn{1}{c}{LaAlO$_3$}&
\multicolumn{1}{c}{$<$$Z_{\rm Al}^*$$>$}&
\multicolumn{1}{c}{$<$$Z_{\rm La}^*$$>$ }&
\multicolumn{1}{c}{$<$$Z_{\rm O}^*$$>$ }&
\multicolumn{1}{c}{$\kappa_{\infty}$}&
\multicolumn{1}{c}{$\kappa_{s}$}\\
    \hline 
  crystal       & 2.96 &   4.37 &   --2.44 & 4.8 & 26.4 \\
 amorphous   & 2.96 &    3.88 &    --2.38 & 4.5 & 27.1 \\
 \hline\hline   \end{tabular}
    \label{tab1}
\end{table}

 In a-LaAlO$_3$, Al has the same non-anomalous dynamical charge as  in the crystal. This is a token of its having no covalent bonds to support a dynamical charge flow upon displacement \cite{zstar} in either case. Al also has a rather crystal-like environment with a coordination of 6 and average angle 90$^{\circ}$ (Fig.\ref{fig1}a and \ref{fig1}b). For La, $Z^*$ is instead severely reduced: the  disruption of ionic-covalent charge flow is clearly due to the major loss of coordination (12 to 6, see Fig.\ref{fig1}c) also signaled by the O-La-O angle distribution in Fig.\ref{fig1}d.
 
In a-DyScO$_3$ the effect is reversed. Dy has essentially the same $Z^*$ as in the crystal, in accordance with its having a similarly well-defined coordination and distance to O (Fig.\ref{fig1}c). Note also the O-Dy-O angle distribution, Fig.\ref{fig1}d, quite less blurred than the O-La-O of a-LaAlO$_3$. The A cation, Sc, suffers instead from an appreciable  $Z^*$ reduction. As we discussed in Ref.\,\onlinecite{dysco}, Sc has a sizably anomalous effective charge in the crystal, as a results of its employing $d$ states in bonding to O. While the Sc coordination is still 6 in a-DyScO$_3$,  the covalent component of the dynamical polarizability is reduced by the strong  octahedra distortion. We remind that the latter shows up in the bimodal angle distribution at 80 and 100$^{\circ}$ (Fig.\ref{fig1}b)  instead of a unimodal around 90$^{\circ}$.
 
In closing this section we  recall our earlier reports \cite{zstar,lalo,dysco} of a sizable reduction of DFT effective-charge anomalies upon use of   self-interaction corrections.\cite{sic} This reduction lead to  total dielectric constants smaller by $\sim$10-20\%, the agreement with experiment being typically improved.  Due to the size of the cells involved here, we did not evaluate the self-interaction-corrected charges in the present cases. The  $\kappa$'s reported here may thus be expected to be somewhat larger than experiment. Of course, the amorphous-crystal comparison is unaffected.

    \begin{figure}[ht]
\centering
\includegraphics[clip,width=4.3cm,height=6.5cm]{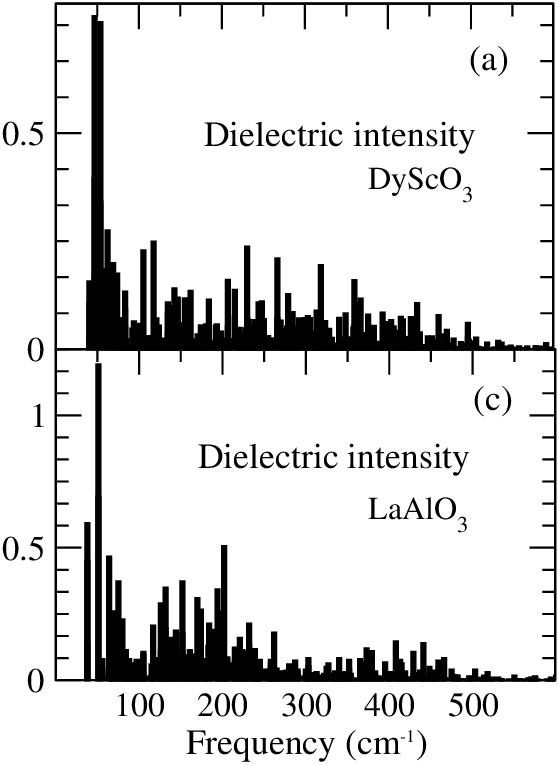}\includegraphics[clip,width=4.2cm]{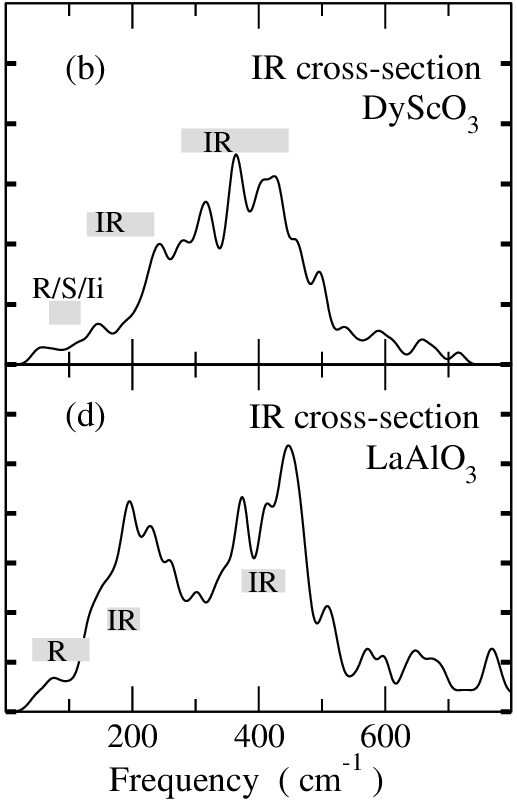}
\caption{\small{Dielectric intensity (DI, left column) and IR spectra (right column) for a-LaAlO$_3$ and a-DyScO$_3$. Panel a:  DyScO$_3$ DI; b: DyScO$_3$ IR; c: LaAlO$_3$ DI;  d: LaAlO$_3$ IR. In panels b and d the approximate positions of crystal vibrational modes \protect{\cite{lalo,dysco}} is indicated. Labels stand for R=Raman, S=silent, Ii=infrared inefficient, IR=infrared. The dielectric intensity is dimensionless (the individual terms in the first equality of Eq.1). The IR spectrum, which has a 10-cm$^{-1}$ lorentzian broadening  built in, is in arbitrary units (same scale for panels b and d).}} \label{fig2}
\end{figure}

\subsection{Dielectric and vibrational properties} In Table \ref{tab1} we reported the average electronic and static dielectric constants. The a-phase static values are clearly in the vicinity of the crystal values, as we originally purported to show. The electronic component is only slightly affected by disorder. Indeed, we find that both a-phases have a crystal-like gap in the electronic density of states, as expected in view of their strong ionicity and of the anion-like and cation-like character of, respectively,  valence and conduction states.  The ionic component, therefore, is also approximately the same  in both phases. To understand why this occurs  despite the reduced effective charges, we  analyze the frequency-dependent dielectric intensity, i.e. the individual (dimensionless) terms in the first equality of Eq.1.

\begin{figure}[ht]
\centering
\includegraphics[clip,width=8.5cm]{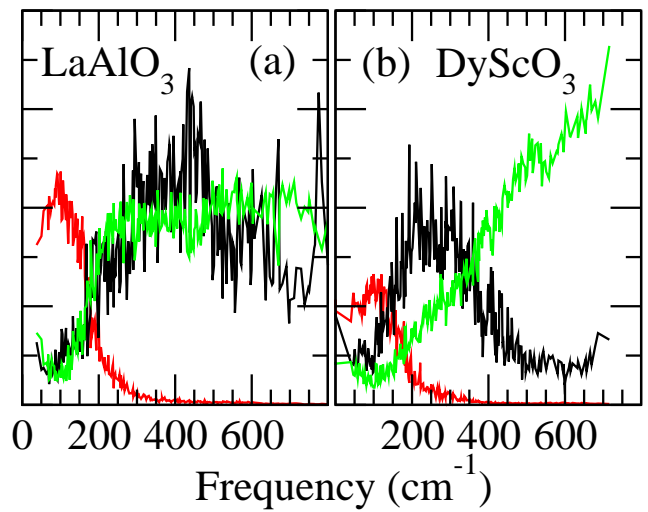}
\caption{\small{(Color on-line) Average atom-specific dipoles  in a-LaAlO$_3$ (panel a) and a-DyScO$_3$ (panel b), i.e. the effective mode charges  $z_{\lambda}$ (Eq.1)  built up as a sum over each atomic species,  normalized to the number of atoms in that species. Black curve: A cations; red (grey): R cation; green (light grey): O. }} \label{fig3}
\end{figure}

 In Fig.\ref{fig2} we report the dielectric intensity (left panels) and IR spectrum (right panels) of a-DyScO$_3$ (top panels) and  a-LaAlO$_3$ (bottom panels).  We indicated  the approximate position of the  calculated Raman, IR, silent modes of the crystals as horizontal bars superposed on the IR spectra of the a-phase. A detailed list is given in Refs.\onlinecite{lalo} and \onlinecite{dysco}.  In Fig.\ref{fig3}, we report the dipole amplitudes decomposed by atomic species, i.e.   the vectors $z_{\lambda}$ as defined in Eq.1, but obtained summing over one species only. 
 
Considering the IR spectrum which, we recall, is a measurable quantity, we  note  that the IR activity at low frequency is  small. However, it is amplified in the dielectric intensity due to the 1/$\omega^2$ factor (Eq.1), so we concentrate on the low frequency region. To anticipate our conclusion, the dielectric enhancement at low frequency is due to IR-inactive (Raman, silent) or inefficient IR crystal modes that acquire IR character or enhance it due to disorder. (Inefficient IR modes in the crystal   are those with weak intensity, say  3 orders of magnitude lower, than typical IR-active crystal modes -- see Ref.\,\onlinecite{dysco}.)

Aside from minor differences of detail, both c-LaAlO$_3$ and c-DyScO$_3$ have their  dominant IR modes  in the range  150-200 cm$^{-1}$. As discussed previously,\cite{lalo,dysco}  these modes derive from  the octahedron-vs-R sublattice vibration typical of perovskites. Further IR modes around 300 to 400 cm$^{-1}$ and above -- relatively  uninteresting for us, being suppressed by the 1/$\omega^2$ factor in Eq.1 -- are intra-octahedron vibrations involving only O and the A cation. These features largely carry over to the a-phases, as  confirmed by the IR spectra (Fig.\ref{fig2}b and d) showing peaks in approximate correspondence with crystal IR activity regions. The  species-projected dipoles (Fig.\ref{fig3}) also confirm this picture : in the 150-200 cm$^{-1}$ region the dipoles generated, respectively, by the R cation and the A-O complex, are similar as expected. At   higher frequency, instead, the displacements are exclusively A and/or O-like, and have intra-octahedron character.

Let us  discuss  the low-frequency dielectric intensity-boosting  modes starting with LaAlO$_3$. We recall that in c-LaAlO$_3$ there exist three Raman  modes (per primitive cell) related to the cubic-rhombohedral instability,\cite{lalo}  at 35 (doubly degenerate) and 130 cm$^{-1}$. They are characterized by octahedra rotations accompanied by non-dipolar R-cation vibrations.  In the disordered environment these modes mix with the IR counter-phase R-(A-O) motions, producing new vibrations with non-zero IR activity. The energies of these mixed modes start at about 40 cm$^{-1}$, and boost the low-frequency  dielectric intensity. The mixed modes can be seen as vibrations of the disordered R sublattice against  A-O units wobbling in a disorderly fashion about their center. This agrees with the dipole at low frequency being mostly contributed by the R cation, as shown in Fig.\ref{fig3}a.

In c-DyScO$_3$, the relevant modes \cite{dysco} are Raman at 100 cm$^{-1}$,   silent at 70 cm$^{-1}$, and  inefficient IR also near 100 cm$^{-1}$. As we have discussed previously in Ref.\,\onlinecite{dysco} -- especially in connection with Fig. 3 thereof -- these three modes have no  dipole (or just  a tiny one) thanks to a delicate symmetry-induced compensation of different displacements. As such, they have a potential for developing some dipole moment in a disordered environment disrupting the symmetric vibrational pattern. We therefore attribute the dielectric intensity boost at low frequency in DyScO$_3$ to the activations of IR activity in these modes. Dy motions are again quite important in this energy region. A peculiarity of DyScO$_3$ is that Sc motions (Fig.\ref{fig3}b) die out at frequencies above 300-400 cm$^{-1}$ (unlike those of Al, Fig.\ref{fig3}a). We note in passing that this is probably a characteristic feature of Sc-O vibrations, as the dipoles are closely analogous to our findings for a-Sc$_2$O$_3$ to be presented elsewhere.

We finally mention the results for the a-LaAlO$_3$  structure generated by ab initio dynamics. The static dielectric constant for this structure is 22.4.  While still in the vicinity of the crystal value, this result is appreciably lower than for the empirical-potential sample ($\sim$27). The decrease is due to a (moderate)  reduction in the  ionic dielectric intensity at low frequency (the electronic part is unchanged, at 4.5). The  reduction should be attributed to the lower  average coordination  of Al sites by about 10\% (5.4 vs 6) in the ab initio sample compared to the empirical-potential sample discussed so. We suspect this to be an artifact caused by  the much higher cooling rate of the  ab initio sample (500 K/psec) compared to the empirical-potential one (50 K/psec) discussed in detail so far. (We recall, anyway,  that all samples were optimized in structure at 0 K after cooling.) 

\section*{Summary and acknowledgments}
In summary,  
we have  shown that the static dielectric response  in amorphous high-$\kappa$  RAO$_3$ oxides  is similar to that of their crystal phase, due to a combination of moderate polarizability reduction and of disorder-induced IR activation of non-polar low-energy modes, as well as to the inheritance by the a-phase of several c-phase structural and vibrational features. We conclude that  the dielectric-constant conservation  upon amorphization of these materials is theoretically plausible. Given their very similar structural and vibrational properties at low frequency\cite{bazro}, we suggest  that  our conclusions may qualitatively apply to perovskites with cations from Group II and IV.

Work partially supported by  EU (project FUNC), MUR Italy  (PON-Cybersar, PRIN05, "Ritorno cervelli"), Fondazione BdS, IMEC industrial affiliation program. Most calculations were done on the high-performance cluster ichnusa@CASPUR Rome.


\begin{thebibliography}{99}

\bibitem{hik}
See MRS Bulletin vol. {\bf 27}, n.3 (2002), Special Issue on Advanced Gate Dielectrics for Microelectronics, and  recent USA press coverage ({\tt\footnotesize www.nytimes.com/2007/01/27/technology/27chip.html}).


\bibitem{lalo}
 P. Delugas, V. Fiorentini,  and A. Filippetti, Phys. Rev. B {\bf 71}, 134302 (2005).


\bibitem{rev}
P. Delugas,   V. Fiorentini, and A. Filippetti, in  {\it Rare Earth Oxide Thin Films: Growth, Characterization and Applications},  M. Fanciulli and G. Scarel eds.  (Springer Topics in Applied Physics, Berlin 2006);
  V. Fiorentini, P. Delugas,  and A. Filippetti, in  {\it Advanced Gate Stacks on High-Mobility Semiconductors}, A. Dimoulas, E. Gusev, P. McIntyre, and M. Heyns eds.  (Springer Series in Advanced  Microelectronics, Berlin, 
 2006).


\bibitem{dysco}
P. Delugas, V. Fiorentini,   A. Filippetti, and G. Pourtois, Phys. Rev. B. {\bf 75}, 115126 (2007).

\bibitem{zstar}
A. Filippetti and N. A. Spaldin, 
Phys. Rev. B {\bf 68}, 045111 (2003).
\bibitem{zhao-zro2}
X. Zhao, D. Ceresoli, and D. Vanderbilt, Phys. Rev. B {\bf 71}, 0085107 (2005)
\bibitem{latio}
I. V. Solovyev, Phys. Rev. B {\bf 74}, 054412 (2006).



\bibitem{lagao}
R. L. Sandstrom, E. A. Giess, W. J. Gallagher, A. SegmŸller, E. I. Cooper, M. F. Chisholm, A. Gupta, S. Shinde, and R. B. Laibowitz, 
Appl. Phys. Lett. {\bf 53},  1874 (1988).

\bibitem{discoamo} 
C. Zhao, T. Witters, B. Brijs, H. Bender, O. Richard, M. Caymax, T. Heeg, 
J. Schubert, V. V. Afanas'ev, A. Stesmans, and D. G. Schlom, 
Appl. Phys.  Lett. {\bf 86}, 132903 (2005);
T. Heeg, M. Wagner, J. Schubert, Ch. Buchal, M. Boese, M. Luysberg, E. Cicerella, and J. L. Freeouf, 
Microelectr. Engin. {\bf 80}, 150 (2005)

\bibitem{laloamo}

L. F. Edge, D. G. Schlom, S. A. Chambers, E. Cicerrella, J. L. Freeouf, B. Holl\"ander, and J. Schubert, 
Appl. Phys. Lett. {\bf 84}, 726 (2004);
X. B. Lu, H. B. Li, Z. H. Chen, X. Zhang, R. Huang, H. W. Zhou, X. P. Wang, B. Y. Nguyen, C. Z. Wang, W. F. Xiang, M. He, and B. L. Cheng, 
Appl. Phys. Lett. {\bf 85}, 3543 (2004).

\bibitem{lalolow}
R. A. B. Devine, J. Appl.   Phys. {\bf 93}, 9938 (2003).

\bibitem{gulp}
J. D. Gale,  Faraday Trans. {\bf 93}, 629 (1997); 
Phil. Mag. B {\bf 73}, 3 (1996).

\bibitem{cool}
K. Vollmayr, W. Kob, and K. Binder, Phys. Rev.         B {\bf 54}, 15808 (1996).

\bibitem{vasp}
G. Kresse and J. Furthm\"uller, Comput. Mater. Sci. {\bf 6}, 15 (1996); Phys. Rev. B {\bf 54}, 11169 (1996); G. Kresse and D. Joubert, {\it ibid.}, {\bf 59}, 1758 (1999). The PAW potentials provided with VASP are used. The GGA is the Perdew  {\it et al.} implementation (see Ref.\,\onlinecite{espresso}).




\bibitem{espresso}
S. Baroni, A. Dal Corso, S. de Gironcoli, P. Giannozzi, C. Cavazzoni, G. Ballabio, S. Scandolo, G. Chiarotti, P. Focher, A. Pasquarello, K. Laasonen, A. Trave, R. Car, N. Marzari, A. Kokalj, 
{\tt http://www.pwscf.org/}.
The GGA approximation by 
J. P. Perdew, J. A. Chevary, S. H. Vosko, K. A. Jackson, M. R. Pederson, D. J. Singh, and C. Fiolhais,  
Phys. Rev. B {\bf 46}, 6671 (1992), and ultrasoft pseudopotential according to  D. Vanderbilt, Phys. Rev. B {\bf 41}, R7892 (1990) are used (see also Refs.\onlinecite{lalo} and \onlinecite{dysco} for technical matters).


\bibitem{dfpt}
S. Baroni, P. Giannozzi, and A. Testa,
Phys. Rev. Lett. {\bf 58}, 1861  (1987);
X. Gonze, Phys. Rev. B {\bf 55}, 10337 (1997); X. Gonze and C. Lee, Phys. Rev. B {\bf 55}, 10355 (1997); S. Baroni, S. de Gironcoli, A. Dal Corso, and P. Giannozzi, Rev. Mod. Phys.  {\bf 73}, 515 (2001).

\bibitem{sic}
A. Filippetti and N. A. Spaldin,
Phys. Rev. B {\bf 67}, 125109 (2003).



%



\bibitem{bazro}
J. W. Bennett, I. Grinberg, and A. M. Rappe, Phys. Rev. B {\bf 73}, 180102R (2006). 
\end{thebibliography}
\end{document}